\documentclass[showkeys,nofootinbib,prd]{revtex4}

\usepackage{amsmath}
\usepackage{amsfonts}
\usepackage{amssymb}
\usepackage{amsthm}
\usepackage{mathtools}
\usepackage{subfigure}
\DeclareFontFamily{U}{mathb}{\hyphenchar\font45}
\DeclareFontShape{U}{mathb}{m}{n}{
      <5> <6> <7> <8> <9> <10> gen * mathb
      <10.95> mathb10 <12> <14.4> <17.28> <20.74> <24.88> mathb12
      }{}
\DeclareSymbolFont{mathb}{U}{mathb}{m}{n}

\DeclareMathSymbol{\Sun}{3}{mathb}{"40}

\allowdisplaybreaks

\begin{document}
\title{Cosmological extra dimensions can mimic dark matter II. Growth of cosmological structures}
\author{Mattia Villani}
\affiliation{University of Urbino Carlo Bo, Department of Pure and Applied Sciences (DiSPeA), Via Santa Chiara, 27, Urbino (PU), 61029, Italy}
\email{mattia.villani@uniurb.it}

\begin{abstract}
This is the second paper of this series in which we prove we can do without dark matter using spacetime with extra dimensions compactified to a hypersphere with radius $R$. Here we prove cosmological structures can grow even in a Universe without dark matter, provided that extra dimensions are contracting.
\end{abstract}
\keywords{Gravity, higher dimensional; Cosmology; Dark matter}
\maketitle

\section{Introduction}

In several papers in the literature (see, for example, \cite{RS1,RS2,RS3,RS4,RS5,RS6,mio, mio3}) authors have proved that using a FLRW spacetime with $n$ extra dimensions compactified to a hypersphere of radius $R$ dependent on time $t$, one can explain the accelerated expansion of the Universe without the need of introducing another form of energy, the dark energy. In the companion paper \cite{mio2}, we have considered the idea that dark matter can be the macroscopic effect of the presence of extra dimensions. There, we have found that we can describe the rotation curve of spiral galaxies using a metric of the form
\begin{equation}
    ds^2=-(1+2\Phi(r))\,dt^2+dr^2+r^2\,d\Omega_2-R(r)^2\,d\Omega_n
\end{equation}
where $\Phi$ is the Newtonian potential and $R$ is the space-dependent radius of the extra dimensions, which are assumed to be compactified to a hypersphere with metric $d\Omega_n$. Here, we continue exploring this idea studying the collapse of cosmological structures in a spacetime with extra dimensions, but without dark matter. In $\Lambda$CDM model, dark matter is fundamental in the collapse of structures, since it provides a potential well where baryonic matter can accumulate. Dark matter is present in a 5:1 ratio with respect to baryonic matter \cite{planck} and while collisionless dark matter eventually virializes, the gas can cool, collapse further, and fragment eventually forming stars and galaxies. Small structures form first and then build larger structures hierarchically up to the scale of galaxy clusters, \cite{review3,form1,form2,form3}.

We shall see that the extra dimensions can drive the collapse of spherical overdensities, leading to a faster growth of structures, even without dark matter. We shall also see that we need contracting extra dimensions in order to have a collapse of matter in the 4d submanifold, consistently to what found in \cite{mio} using Supernova data.  Unfortunately, our model does not have analytical solutions, so we need to resort to numerical analysis.

The plan of the paper is as follows: in Section \ref{sec:back}, we introduce the Friedman equations derived in \cite{mio}, needed in this work; in Section \ref{sec:linear}, we study the linear growth of structures, while in Section \ref{sec:nlin} the non-linear growth. Finally, in Section \ref{sec:concl} we conclude our exposition.

\section{The background cosmological model}
\label{sec:back}

The background cosmological model was described in literature, for example in \cite{RS1,RS2,RS3,RS4,RS5,RS6}. The metric considered has the form
\begin{equation}
    ds^2=dt^2-\dfrac{a(t)^2}{1-r^2\,\kappa}dr^2-a(t)^2\,r^2\,d\Omega_2-R(t)^2\,d\Omega_n
\end{equation}
The Einstein equations can be reduced to three equations (in the case of perfect fluid with density $\rho$ and pressure $p$; we assume that the stress energy tensor is zero in the extra dimensions, i.e, that matter is confined to the 4d submanifold)
\begin{subequations}
    \begin{equation}\label{eq:fr1}
        \dfrac{\kappa}{a^2}+\left(\dfrac{\dot{a}}{a}\right)^2+n\,\dfrac{\dot{a}}{a}\dfrac{\dot{R}}{R}+\dfrac{n(n+1)}{6}\,\left[\dfrac{1}{R^2}+\left(\dfrac{\dot{R}}{R}\right)^2\right]=\dfrac{8\pi G \,\rho}{3},
    \end{equation}
    \begin{equation}\label{eq:fr2}
        \dfrac{\ddot{a}}{a}+n\dfrac{\dot{a}}{a}\,\dfrac{\dot{R}}{R}+\dfrac{n(n-1)}{3}\,\left[ \dfrac{1}{R^2}+\left(\dfrac{\dot{R}}{R}\right)^2 \right]=-\dfrac{4\,\pi\,G\,\rho}{3}\,\left(\dfrac{2\,(1+2n)}{(n+2)}\,\rho-3p\right)
    \end{equation}
    \begin{equation}\label{eq:fr3}
    \begin{split}
        \dfrac{\ddot{R}}{R}=(1-n)\,\left[ \dfrac{1}{R^2}+\left( \dfrac{\dot{R}}{R} \right)^2 \right]-3\,\dfrac{\dot{a}}{a}\dfrac{\dot{R}}{R}+\dfrac{4\pi\,G\,\rho}{n+2}
    \end{split}
    \end{equation}
\end{subequations}
The first two are the usual equations modified by the presence of the extra dimensions, while the third is the new equation describing the evolution of the radius $R$. In the following, we assume $\kappa=0$, a spatially flat spacetime.

We were able to find two FLRW-like solutions to the above equations for the case of flat spacetime. In order to find them, we eliminate from \eqref{eq:fr1} the term $R^{-2}$ using \eqref{eq:fr3}, obtaining (neglecting the pressure)
\begin{equation}\label{eq:fr4}
    \left( \dfrac{\dot{a}}{a} \right)^2+n\,\dfrac{\dot{a}}{a}\dfrac{\dot{R}}{R}-\dfrac{n}{6}\,\dfrac{\ddot{R}}{R}=\dfrac{4\pi\,G\,\rho}{3}\,\dfrac{4+n}{2+n}.
\end{equation}
This has the simple power-law solution
\begin{equation}
    R=\pm|c|\,t, \qquad a=b\,t^s, \qquad c\neq0
\end{equation}
with
\begin{align}\label{eq:dust1}
    &s=\dfrac{2}{3},\quad b=\left(\dfrac{24\,\pi\,G\,\rho_0\,(4+n)}{16+20n+6n^2}\right)^{1/3} & \rho=\dfrac{\rho_0}{a^3},\\
    &s=\dfrac{1}{2},\quad b=\left( \dfrac{2}{3}\,\dfrac{8\pi\,G\,\rho_0\,(4+n)}{1+3n+n^2} \right)^{1/4}& \rho=\dfrac{\rho_0}{a^4}.
\end{align}

In fact, this is the simplest solution of the family 
\begin{equation}
    R=c\,t^m, \qquad a=b\,t^s, \qquad c\neq0
\end{equation}
with
\begin{align}
    &s=\dfrac{2}{3},\quad b=\left(\dfrac{24\pi\,G\,\rho_0\,(n+4)}{16+8n+18mn-6m^2n+9mn^2-3m^2n^2}\right)^{1/3} & \rho=\dfrac{\rho_0}{a^3},\\
    &s=\dfrac{1}{2},\quad b=\left( \dfrac{18\pi\,G\,\rho_0\,(4+n)}{(6+3n+10mn-4m^2n-5mn^2-2m^2n^2)} \right)^{1/4}& \rho=\dfrac{\rho_0}{a^4}.
\end{align}
We notice that the constant solution $R=c$ is also a solution to equation \eqref{eq:fr4}. This result suggests that probably one can find FLRW-like solutions even for non flat geometries.

In this paper we shall focus on the first solution \eqref{eq:dust1}, but the equations and the results can be easily generalized.

\section{Linear growth}
\label{sec:linear}
\subsection{The equation for overdensities}
In this Section, we consider the linear perturbation theory. We consider a metric of the form (see \cite{mio2})
\begin{equation}\label{eq:met}
    ds^2=-(1+2\Phi(r))\,dt^2+dr^2+r^2\,d\Omega_2-R(t,r)^2\,d\Omega_n
\end{equation}
where $d\Omega_i$ is the metric of a $i$-dimensional sphere, $R$ is the radius of the extra dimensions and \begin{equation}
    \Phi=-\dfrac{1}{2}\,\left( \dfrac{\ddot{a}}{a} \right)\,|x|^2, \qquad |x|=a\,|r|
\end{equation}
where $|x|$ is the comoving distance. Above we have imposed $c=1$.

The  continuity and Euler equations are derived from $T^{\mu}_{\phantom{\mu}\nu;\mu}=0$, and are given by
\begin{subequations}
    \begin{equation}
        \dot{\rho}+3\,H\,(p+\rho)=0,
    \end{equation}
    \begin{equation}
        \dot{u}_\mu=-\dfrac{h^\nu_{\phantom{\mu}\nu}\,\nabla_\nu\,p}{(\rho+p)}
    \end{equation}
\end{subequations}
where $h_{\mu\nu}$ is the spatial part of the background metric and $u^\mu$ the velocity of fluid (we recall that we assume there is no motion of matter in the extra dimensions, thus $u^\mu=0$ for $\mu>3$). Following \cite{pad}, we can rewrite the Euler equation as
\begin{equation}
    \dot{H}+H^2+\dfrac{4\pi G}{3}\,(\rho+3p)+\dfrac{1}{4}\,\dfrac{\nabla p}{p+\rho}=0
\end{equation}
We have 
\begin{equation}
    \dfrac{d\tau}{dt}=1-\dfrac{\delta p}{\rho+p}
\end{equation}
then, considering $H=H_b+\delta H$, where $H_b$ is the background Hubble parameter and $\delta H$ its perturbations, we find at the leading order
\begin{subequations}
    \begin{equation}\label{eq:first}
        \dot{H}_b+H_b^2+\dfrac{4\pi G}{3}\,(\rho_b+3p_b)+\dfrac{1}{4}\,\dfrac{\nabla p_b}{p_b+\rho_b}=0,
    \end{equation}
while at first order we have
    \begin{equation}\label{eq:hub_pert}
        \dot{\delta H}+2\,H_b\,\delta H+\dfrac{4\pi G \delta\rho}{3}+\dfrac{1}{3}\,\dfrac{\nabla \delta p}{\rho+p}-\dfrac{\delta p}{p+\rho}\,\left( \dfrac{1}{3}\,\dfrac{1}{R_b^2}+2\,\mathcal{H}_b+\mathcal{H}_b^2 \right)=0;
    \end{equation}
\end{subequations}
we notice that extra dimensions introduce a correction, the last term in the parentheses, which is absent in the usual model. $R_b$ and $\mathcal{H}_b$ are the background radius and expansion rate of the extra dimensions; there are perturbations to these parameters, as we shall show below. The continuity equation becomes
\begin{subequations}
    \begin{equation}
        \dot{\rho}_b+3\,H_b\,(p_b+\rho_b)=0,
    \end{equation}
at the leading order, while at first order we have
    \begin{equation}\label{eq:cont_per}
        \dot{\delta\rho}+3\,(\rho_b+p_b)\,\delta H+3\,H_b\,\,\delta\rho=0.
    \end{equation}
\end{subequations}
Now we define
\begin{equation}
    \delta=\dfrac{\delta\rho}{\rho}, \qquad w=\dfrac{p}{\rho}, \qquad v^2=\dfrac{\delta p}{\delta\rho},
\end{equation}
and rewrite \eqref{eq:cont_per} as follows
\begin{equation}\label{eq:dh}
    \delta H=-\dfrac{1}{3(1+w)}\,\left[ \dot{\delta}-3\,H_b\,w\,\delta \right].
\end{equation}
We substitute this into \eqref{eq:hub_pert}, using that
\begin{equation}
    \dot{w}=3\,H_b\,(w-v^2)\,(1+w), \qquad \dot{H}_b=-\dfrac{3}{2}\,H_b^2\,(1+w),
\end{equation}
thus obtaining an equation for the evolution of the overdensities $\delta$
\begin{equation}
\begin{split}
    &\ddot{\delta}-(6w-3v^2-2)\,H_b\,\dot{\delta}-\dfrac{3}{2}\,\Big[ H_b^2\,(1-6v^2+8w-3w^2) +\\
    &- \left( 1+w-2v^2 \right)\,\left( \dfrac{n(n-1)}{6}\,\dfrac{1}{R_b^2}+2\,n\,\mathcal{H}_b+\dfrac{n(n-1)}{6}\,\mathcal{H}_b^2 \right) \Big]\,\delta=v^2\,\nabla^2\delta,
\end{split}
\end{equation}
which reduces to the usual one when $n\rightarrow0$. If we change the independent variable $t\rightarrow a$ and use a Fourier transform in the term in the right hand side, we find
\begin{equation}\label{eq:dens}
\begin{split}
    a^2\,H_b^2\,{\delta}^{\prime\prime}&-a\,(5w-2v^2-1)\,H_b^2\,{\delta}^\prime-\dfrac{3}{2}\,\Bigg[ H_b^2\,(1-6v^2+8w-3w^2)-\dfrac{2}{3}\, \dfrac{k^2\,v^2}{a^2}+\\
    &+ \left( 1+w-2v^2 \right)\,\left( \dfrac{n(n-1)}{6}\,\dfrac{1}{R_b^2}+2\,n\,a\,H_b^2\,\mathcal{H}_b+a^2\,\dfrac{n(n-1)}{6}\,H_b^2\,\mathcal{H}_b^2  \right) \Bigg]\,\delta=0,
\end{split}
\end{equation}
where a prime indicates a derivative with respect to $a$. The term in the second line is the additional term which comes from the dynamics of the extra dimensions; in the parentheses in the second line, factors $a$ and $a^2$ have appeared because we have transformed the time derivative $\dot{R}/R$ into an $a$-derivative. Finally, we consider the special cases of radiation $w=v^2=1/3$ and dust $w=v^2=0$. In the case of dust, we keep the term proportional to $k^2$. The final result for radiation is:
\begin{subequations}
\begin{equation}\label{eq:rad}
    H_b^2\,a^2\,{\delta}^{\prime\prime}-\left[ 2\,H_b^2- \left( \dfrac{n(n-1)}{6}\,\dfrac{1}{R_b^2}+2\,n\,a\,H_b^2\,\mathcal{H}_b+\dfrac{n(n-1)}{6}\,a^2\,H_b^2\,\mathcal{H}_b^2 \right)-\dfrac{k^2v^2}{3a^2} \right]\,\delta=0,
\end{equation}
while for dust we find
\begin{equation}\label{eq:dust}
    H_b^2\,a^2\,{\delta}^{\prime\prime}+\dfrac{3}{2}\,H_b^2\,{\delta}^\prime-\dfrac{3}{2}\,\left[ H_b^2-\left( \dfrac{n(n-1)}{6}\,\dfrac{1}{R_b^2}+2\,n\,a\,H_b^2\,\mathcal{H}_b+\dfrac{n(n-1)}{6}\,a^2\,H_b^2\,\mathcal{H}_b^2 \right)-\dfrac{2}{3}\,\dfrac{k^2v^2}{a^2} \right]\,\delta=0.
\end{equation}
\end{subequations}
Equations \eqref{eq:rad} and \eqref{eq:dust} are the equations that describe the evolution of the overdensities; both reduce to the usual ones when $n\rightarrow0$; we see that in our model they are coupled with the evolution of the extra dimensions, so we need to derive equations for them as well.

\subsection{The equations for extra dimensions}
In order to obtain the perturbation of the extra dimensions radius, we start from equation \eqref{eq:fr3}, write $R=R_b+\delta R$ and separate the orders, thus obtaining
\begin{subequations}
    \begin{equation}
        \ddot{R}_b=R_b\,\left[ -3\,n H_B\,\dfrac{\dot{R}_b}{R_b}+(1-n)\,\left(\dfrac{\dot{R}_b}{R_b}\right)^2+\dfrac{1-n}{R_b^2} + 2\pi G\,\rho_b\right],
    \end{equation}
    \begin{equation}
    \begin{split}\label{eq:sec}
        \ddot{\delta R}&=\delta R\,\left[ -3 H_B\,\dfrac{\dot{R}_b}{R_b}+(1-n)\,\left(\dfrac{\dot{R}_b}{R_b}\right)^2+\dfrac{1-n}{R_b^2} + 3\pi G\,\rho_b\right]+\\
        &+ R_b\,\left[ -3\,\delta H \,\dfrac{\dot{R}_b}{R_b} +3\,H_b\,\left( \dfrac{\dot{\delta R}}{R_b}-\dfrac{\dot{R}_b}{R_b}\,\dfrac{\delta R}{R_b} \right)+2\,(1-n)\,\dfrac{\dot{R}_b}{R_b}\,\left( \dfrac{\dot{\delta R}}{R_b}-\dfrac{\delta R}
        {R_b} \right) +2\pi G \delta\rho \right],
    \end{split}
    \end{equation}
\end{subequations}
respectively for the leading order and the first order perturbation. The second equation shows that overdensities will source local changes in the radius of the extra dimensions; this gives a physical basis to our hypothesis in the companion paper \cite{mio2}, that locally $R=R(r)$. The second equation is the source of the (fake) dark matter halos. For the moment, we focus on the first, where we change the variable from $t$ to $a$:
\begin{equation}
    \ddot{R}_b=\left(\dot{H}_b+H_b^2\right)\,a\,R^\prime_b+H_b^2\,a^2\,R_b^{\prime\prime}=R_b\,\left[ -3 H_b^2\,a\,\dfrac{R^\prime_b}{R_b}+(1-n)\,H_b^2\,a^2\,\left(\dfrac{R^\prime_b}{R_b}\right)^2+\dfrac{1-n}{R_b^2} + 2\pi G\,\rho_b\right],
\end{equation}
where a prime stand for a derivative with respect to $a$. From equation \eqref{eq:first}, we obtain
\begin{equation}
\begin{split}
    H_b^2\,a^2\,R^{\prime\prime}_b&=-\left( \dfrac{4\pi G}{3}\,(\rho_b+3p_b)+\dfrac{1}{4}\,\dfrac{\nabla p_b}{p_b+\rho_b} \right)\,a\,R^\prime_b+R_b\,\left[ -3 H_b^2\,a\,\dfrac{R^\prime_b}{R_b}+(1-n)\,H_b^2\,a^2\,\left(\dfrac{R^\prime_b}{R_b}\right)^2+\dfrac{1-n}{R_b^2} + 3\pi G\,\rho_b\right]=\\
    &=-\left( \dfrac{4\pi G\,\rho_b}{3}\,(1+3w)+\dfrac{1}{4}\,\dfrac{v^2\,\nabla \rho_b}{\rho_b\,(1+w)} \right)\,a\,R^\prime_b+R_b\,\left[ -3 H_b^2\,a\,\dfrac{R^\prime_b}{R_b}+(1-n)H_b^2\,a^2\,\left(\dfrac{R^\prime_b}{R_b}\right)^2+\dfrac{1-n}{R_b^2} + 2\pi G\,\rho_b\right].
\end{split}
\end{equation}
Now, we specialize the above equation for radiation and dust, obtaining, respectively (we assume that the background density $\rho_b$ is isotropic, so $\nabla \rho_b=0$)
\begin{subequations}
    \begin{equation}\label{eq:rad2}
        H_b^2\,a^2\,R^{\prime\prime}_b=-\left( \dfrac{8\pi G\,\rho_b}{3} \right)\,a\,R^\prime_b+R_b\,\left[ -3 H_b^2\,a\,\dfrac{R^\prime_b}{R_b}+(1-n)\,H_b^2\,a^2\,\left(\dfrac{R^\prime_b}{R_b}\right)^2+\dfrac{1-n}{R_b^2} + 2\pi G\,\rho_b\right],
    \end{equation}
    \begin{equation}\label{eq:dust2}
        H_b^2\,a^2\,R^{\prime\prime}_b=-\left( \dfrac{4\pi G\,\rho_b}{3} \right)\,a\,R^\prime_b+R_b\,\left[ -3 H_b^2\,a\,\dfrac{R^\prime_b}{R_b}+(1-n)\,H_b^2\,a^2\,\left(\dfrac{R^\prime_b}{R_b}\right)^2+\dfrac{1-n}{R_b^2} + 2\pi G\,\rho_b\right].
    \end{equation}
\end{subequations}

From equation \eqref{eq:sec} which describes the linear growth of the perturbations of the radius $R$, we get, changing again the variable to $a$
\begin{equation}
\begin{split}
    \delta R^{\prime\prime}\,H_b^2\,a^2&=\left( \dfrac{4\pi G\,\rho_b}{3} +3\,H_b^2\,a+2(1-n)\,\dfrac{R'_b}{R_b}\,H_b^2\,a^2 \right)\,\delta R^\prime+\\
    &+\delta R\,\left[ -(9-n) H_b^2\,a\,\dfrac{{R}^\prime_b}{R_b}+(1-n)\,\left(H_b\,a\,\dfrac{{R}^\prime_b}{R_b}\right)^2+\dfrac{1-n}{R_b^2} + 3\pi G\,\rho_b\right]+\\
        &+R_b\,\Bigg[ -3\,\dfrac{R^\prime_b}{R_b}\,H_b\,a\, \delta\,H +2\pi G \delta\,\rho_b\Bigg]
\end{split}
\end{equation}
for dust, while for radiation, we get
\begin{equation}
\begin{split}
    \delta R^{\prime\prime}\,H_b^2\,a^2&=\left( \dfrac{8\pi G\,\rho_b}{3} +3\,H_b^2\,a+2(1-n)\,\dfrac{R'_b}{R_b}\,H_b^2\,a^2 \right)\,\delta R^\prime+\\
    &+\delta R\,\left[ -(9-n) H_b^2\,a\,\dfrac{{R}^\prime_b}{R_b}+(1-n)\,\left(H_b\,a\,\dfrac{{R}^\prime_b}{R_b}\right)^2+\dfrac{1-n}{R_b^2} + 3\pi G\,\rho_b\right]+\\
        &+R_b\,\Bigg[ -3\,\dfrac{R^\prime_b}{R_b}\,H_b\,a\, \delta\,H +2\pi G \delta\,\rho_b\Bigg],
\end{split}
\end{equation}
where $\delta H$ is given by equation \eqref{eq:dh} which contains $\delta^\prime$. As said above, these equations describe the linear growth of the (fake) dark matter halos and as can be seen they are tightly coupled to the growth of the overdensities, thus, as found in \cite{mio2}, in this formalism, the halos are sourced by the overdensities.

We now study the evolution of the system.

\subsection{The linear evolution}

In order to obtain the evolution of the overdensities, we need to couple equations \eqref{eq:rad} and \eqref{eq:rad2} for the radiation domination phase and \eqref{eq:dust} and \eqref{eq:dust2} for the matter domination phase. We need an expression for the background Hubble parameter. Changing variable in equation \eqref{eq:fr1}, going from $t$ to $a$, we obtain
\begin{equation}
    H_b^2+n\,H_b^2\,a\,\dfrac{R^\prime_b}{R_b}+\dfrac{n(n-1)}{6R^2_b}+\dfrac{n(n-1)\,H_b^2\,a^2}{6}\,\left( \dfrac{R^\prime_b}{R_b} \right)^2=\dfrac{8\pi G \rho_b}{3},
\end{equation}
from which we derive
\begin{equation}
    H_b^2=\left( \dfrac{8\pi G \rho_b}{3} - \dfrac{n(n-1)}{6R^2_b} \right)\,\left[ 1+n\,a\,\dfrac{R^\prime_b}{R_b}+\dfrac{a^2\,n(n-1)}{3}\,\left( \dfrac{R^\prime_b}{R_b} \right)^2 \right]^{-1},
\end{equation}
which reduces to the usual expression when there are no extra dimensions.

\begin{figure}[htb]
    \centering
    \includegraphics[width=0.65\linewidth]{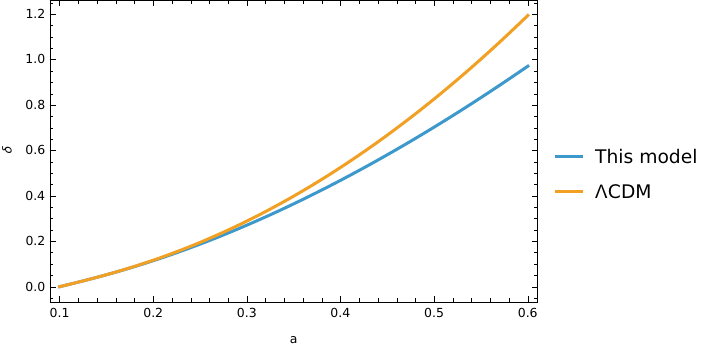}
    \caption{Linear evolution of the overdensities in radiation domination epoch for $k\ll1$ and $n=2$.}
    \label{fig:delta_rad}
\end{figure}

\begin{figure}[htb]
    \centering
    \includegraphics[width=0.65\linewidth]{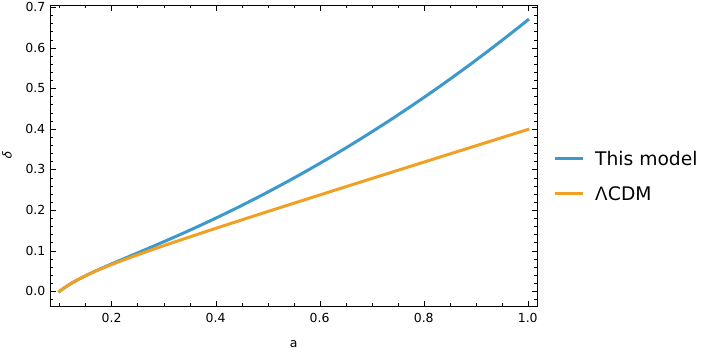}
    \caption{Linear evolution of the overdensities in matter domination epoch for $k\ll1$ and $n=2$.}
    \label{fig:delta_mat}
\end{figure}

As in \cite{pad}, we need to distinguish the case $k\ll1$ from the case $k\gg1$. The results for the first case are reported in Figures \ref{fig:delta_rad} and \ref{fig:delta_mat}, respectively, where the blue curves represent our model and the yellow curves the usual $\Lambda$CDM model. In both cases we plot the results for $n=2$, as an example; the effects of $n$ are studied below.  We see that in the radiation domination phase $\Lambda$CDM overdensities grow faster than in our model: they become $O(1)$ before our model; in fact $\delta_{\Lambda CDM}\propto a^2$, while $\delta_{\text{this work}}\propto a^{1.67}$. On the other hand, during the matter domination phase, our model grows faster than $\Lambda$CDM model, $\delta_{\Lambda CDM}\propto a$, while $\delta_{\text{this work}}\propto a^{1.87}$ and becomes non-linear much sooner than the $\Lambda$CDM model.

\begin{figure}[htb]
    \centering
    \includegraphics[width=0.70\linewidth]{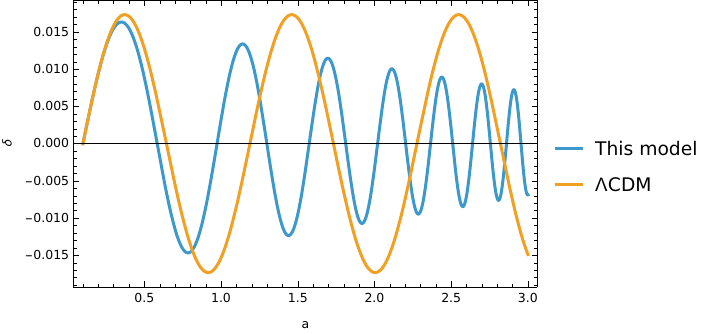}
    \caption{Linear evolution of the overdensities in radiation domination epoch for $k\gg1$ and $n=2$.}
    \label{fig:delta_rad_large}
\end{figure}

The second case, in which $k\gg a^2H_b^2$, $n=2$, which means that the wavelength associated to the overdensity is smaller than the Hubble radius ($\lambda\ll d_H$), is reported in Figures \ref{fig:delta_rad_large} for the radiation domination epoch. We see that we have damped oscillations. The oscillations mean that the average overdensity is zero and that the amplitude decreases with time, so the smallest overdensities tend to be washed out during the radiation domination phase and only the largest survive; this is at variance with respect to the $\Lambda$CDM case, where the overdensities are not washed out at any scale.


\subsection{The effect of the number of the extra dimensions}

\begin{figure}
    \centering
    \subfigure{\includegraphics[width=0.6\linewidth]{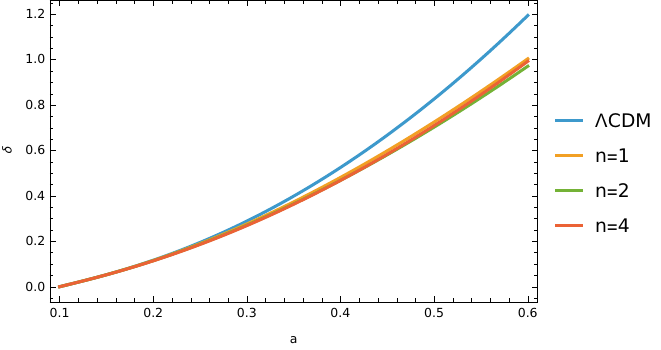}}
    \subfigure{\includegraphics[width=0.6\linewidth]{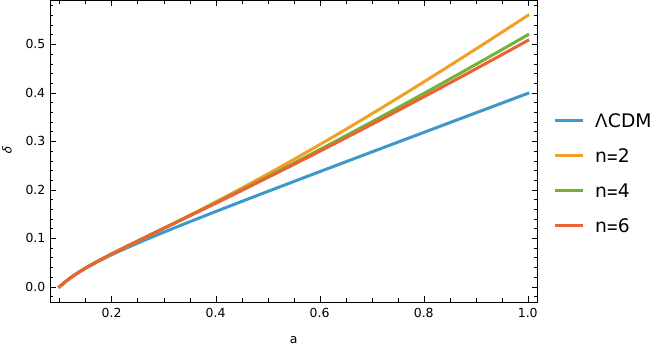}}
    \subfigure{\includegraphics[width=0.6\linewidth]{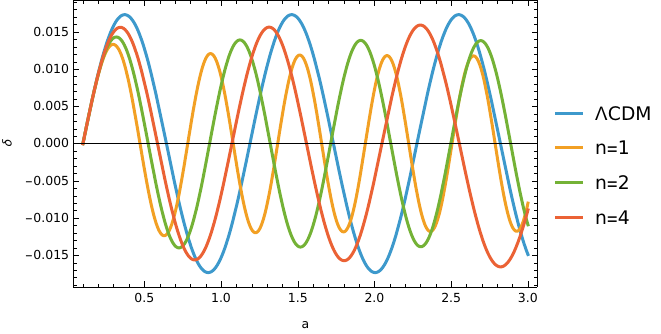}}
    \caption{The effect of the number of extra dimensions. Top panel: radiation domination epoch; central panel: matter domination epoch; bottom panel: the case $k\gg1$ for the radiation domination epoch.}
    \label{fig:lin_dim}
\end{figure}

Here, we vary the number of extra dimensions to see their effect on the linear growth of overdensities. We report our results in Figure \ref{fig:lin_dim}. In the top panel, we report the growth of structures in the radiation domination epoch for $k\ll1$; in the central panel the case of matter domination epoch, again for $k\ll1$; finally, in the bottom panel, we report the case of $k\gg1$ for the radiation domination epoch. We consider $1\leq n\leq 6$, but we see that the effect on the the growth of structure is limited, especially in the top panel; we notice that as $n$ grows the evolution of the strctures get slowly closer and closer the  $\Lambda$CDM model. From the bottom panel, we see that the number of extra dimension changes the frequency and the amplitude of the oscillations and that for large $n$ the damping we have observed in the case $n=2$ in Figure \ref{fig:delta_rad_large} is limited.

\section{The non-linear evolution}
\label{sec:nlin}
The growing overdensities will reach a point in which the linear description given above will no longer be applicable, thus a full non-linear treatment is in order. As in \cite{pad}, we consider the evolution after the overdensities enter the Hubble radius; thus a Newtonian approximation can be used. As in \cite{pad}, we consider a spherical region of radius $r$ containing an overdensity. The Newtonian potential is given by
\begin{equation}
    \begin{split}
        \Phi&=\Phi_b+\delta\Phi=-\dfrac{1}{2}\,\left( \dfrac{\ddot{a}}{a} \right)\,r^2+\delta\Phi=\\
        &=-\dfrac{1}{2}\,\left[ -\dfrac{4(1+2n)\pi G\,\rho_b}{n+2}-n\,\left( \dfrac{\dot{a}}{a} \right)\left( \dfrac{\dot{R}}{R} \right)-\dfrac{n(n-1)}{3}\,\left( \dfrac{1}{R^2}+\left( \dfrac{\dot{R}}{R} \right)^2 \right) \right]\,r^2+\delta\Phi.
    \end{split}
\end{equation}
The potential in $\Lambda$CDM model is given by
\begin{equation}
    \Phi=\dfrac{2\pi G\,\rho_b}{3}\,r^2+\delta\phi,
\end{equation}
so we see that in our model there are several modifications: first the coefficient in front of the background density is different, but there are also new additional terms coming from the extra dimensions dynamics. The acceleration is given by
\begin{equation}
\begin{split}\label{eq:nonlin}
    \dfrac{d^2\vec{r}}{dt^2}=-\dfrac{2(1+2n)}{n+2}\,\dfrac{GM}{r^3}\,\vec{r}-n\,\left[ \left( \dfrac{\dot{a}}{a} \right)\left( \dfrac{\dot{R}}{R} \right)+\dfrac{n(n-1)}{3}\,\left( \dfrac{1}{R^2}+\left( \dfrac{\dot{R}}{R} \right)^2 \right) \right]\,\vec{r},
\end{split}
\end{equation}
where $M_b$ is the background mass enclosed within the sphere and where we have defined the total mass $M$. We see that the equation is different from the usual one, not only because of the coefficient in front of the Newtonian force, but also for the presence of a radial external force depending on the dynamics of the extra dimensions. This force can promote or counteract the collapse depending on the sign of $\dot{R}$: in general, if $\dot{R}\lesssim0$ (collapsing extra dimensions), it increases the acceleration; on the other hand, if $\dot{R}\gtrsim0$ (expanding extra dimensions) it reduces the acceleration, counteracting the collapse. We notice that our equation reduces to the usual one when $n\rightarrow0.$

\begin{figure}[htb]
    \centering
    \subfigure{\includegraphics[width=0.6\linewidth]{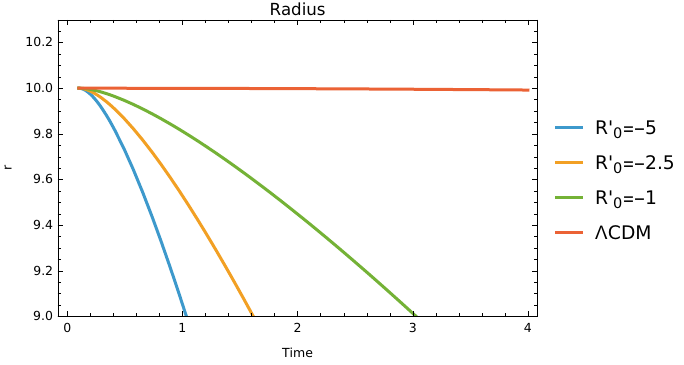}} \subfigure{\includegraphics[width=0.6\linewidth]{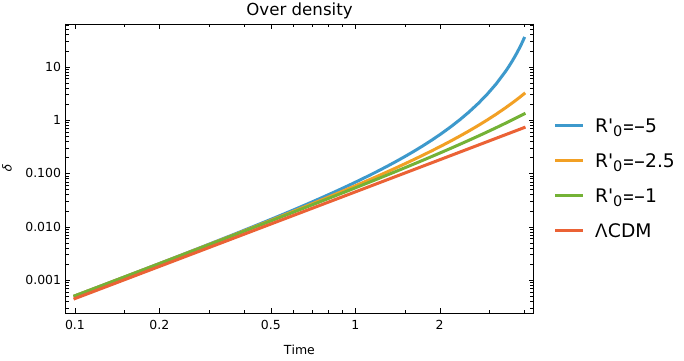}} \subfigure{\includegraphics[width=0.6\linewidth]{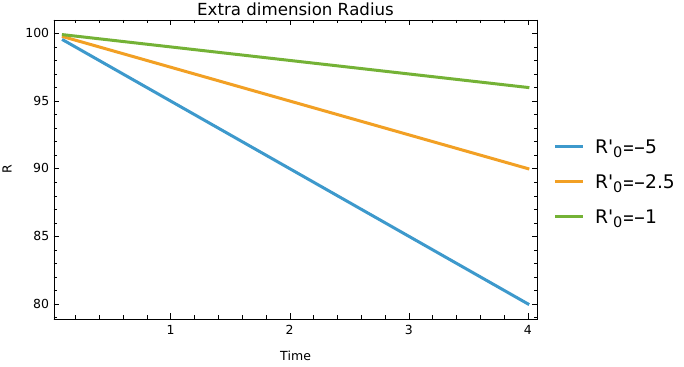}}
    \caption{First panel: evolution of the radius of the shell; Second panel: evolution of the overdensities. Bottom panel: evolution of the radius of the extra dimensions. The blue lines represent the evolution for a $R'(0)=-1$; the yellow lines for $R'(0)=-1/2$; the green lines for $R'(0)=0$. The red lines are for the evolution in the $\Lambda$CDM model. We assume that there are 2 extra dimensions.}
    \label{fig:contr}
\end{figure}

\begin{figure}[htb]
    \centering
    \subfigure{\includegraphics[width=0.6\linewidth]{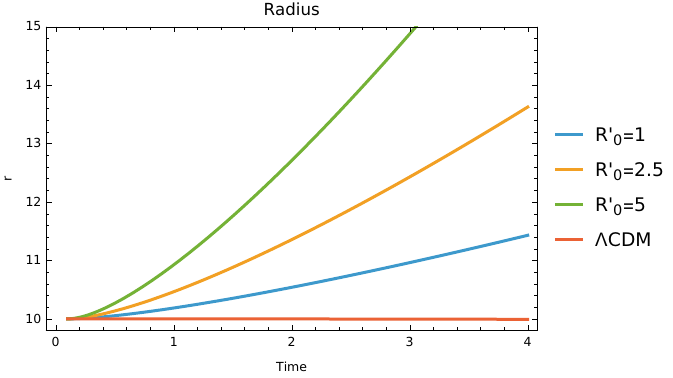}} \subfigure{\includegraphics[width=0.6\linewidth]{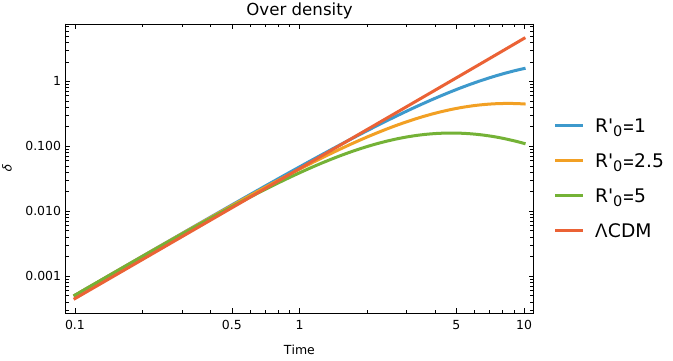}} \subfigure{\includegraphics[width=0.6\linewidth]{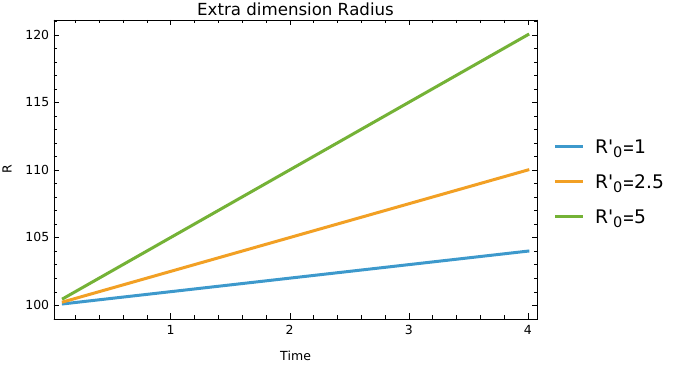}}
    \caption{First panel: evolution of the radius of the shell; Second panel: evolution of the overdensities. Bottom panel: evolution of the radius of the extra dimensions. The blue lines represent the evolution for a $R'(0)=2$; the yellow lines for $R'(0)=1$; the green lines for $R'(0)=1/2$. The red lines are for the evolution in the $\Lambda$CDM model. We assume that there are 2 extra dimensions.}
    \label{fig:exp}
\end{figure}

We study the evolution of the overdensities using the FLRW-like solution \eqref{eq:dust1}. We report the result of the analysis of these equations in Figure \ref{fig:contr}, for the case in which the extra dimensions are contracting and in Figure \ref{fig:exp} for the case in which they are expanding. In both Figures there are three panels: at the top we report the evolution of the radius $r$; in the second, the evolution of the overdensity $\delta=\rho/\rho_b$;  in the bottom panel we report the radius $R$ of the extra dimensions. We notice that we imposed a somewhat large value for $R(0)$: this is due to the fact that we also consider collapsing extra dimensions: if $R(0)$ is \emph{small}, we reach $R=0$ sooner, and a negative radius has no physical meaning.

Starting from Figure \ref{fig:contr}, we notice that in the case of contracting extra dimensions, the radius $r$ reduces much faster than in $\Lambda$CDM model (the red line); at the same time we see that the overdensities grow much faster than usual, we also notice that the larger $R^\prime(0)$, the sooner the overdensities become non-linear; this usually happens when the $\Lambda$CDM model predicts that the overdensities are still into the linear regime. This is due to the fact that $r$ decreases faster in our model.

If we now look at Figure \ref{fig:exp} for the case of expanding extra dimensions, we notice that for the considered values of $R^\prime(0)$, the extra term in equation \eqref{eq:nonlin} leads to an expansion of the radius $r$, thus the overdensity is not actually contracting. In the second panel, we see, however, that the overdensities grow at the beginning at the same rate as in the $\Lambda$CDM model, however as time passes the growth is stopped and the overdensities start decreasing; this happens sooner for larger values of $R'(0)$. This is due the fact that the radius $r$ is not contracting: at the beginning the overdensities grow because the background density decreases, but after some time the increase of $r$ will lead to a decrease of $\delta$. This suggests that the growth of overdensities require collapsing extra dimensions.

In the case of contracting extra dimensions, we can have a contraction of the overdense region which is faster than in $\Lambda$CDM means \emph{in principle} that we can do without DM at all, because smaller clumps of matter will collapse at the same rate as in $\Lambda$CDM model; however, the collapse of baryonic matter is more complicated than this: baryonic matter will warm during collapse and if the heat is not radiated away efficiently, the collapse will be slowed down, or even stopped. The potential well of the DM helps the collapse even for \emph{small} overdensities \cite{pad}. Here, we do not have this potential well, so dedicated numerical studies of the evolution of the overdensities taking into consideration radiation processes must be performed in order to estimate the minimum mass that the overdensities should have for the collapse to be successful (Silk mass).

\begin{figure}[htb]
    \centering
    \subfigure{\includegraphics[width=0.6\linewidth]{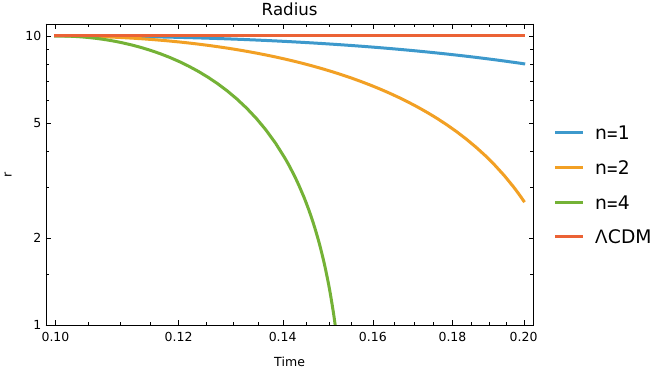}} \subfigure{\includegraphics[width=0.6\linewidth]{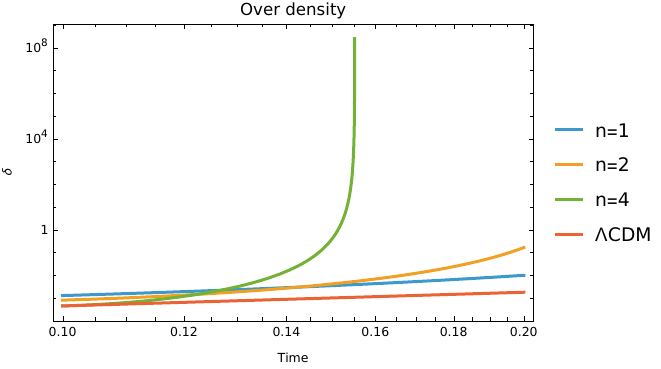}}
    \caption{Top panel: evolution of the radius of the shell as a function of the number of extra dimensions; Bottom panel: evolution of the overdensities. The blue lines represent the evolution for a $n=1$; the yellow lines for $n=2$; the green lines for $n=4$. The red lines are for the evolution in the $\Lambda$CDM model. We assume that $R'(0)=-1$.}
    \label{fig:dim}
\end{figure}

Finally, in Figure \ref{fig:dim}, we report the evolution of the radius (top panel) and of the overdensities (bottom panel) as a function of the number of dimensions. We see that the larger the number of extra dimensions, the faster the overdensity collapses. In the case $n=4$ the overdensities become strongly non-linear, while the $\Lambda$CDM model predicts that they are still well into the linear regime.

\section{Conclusions and discussion}
\label{sec:concl}

In this paper and in its companion \cite{mio2} we have followed the hypothesis that we can do without dark matter if we introduce cosmological extra dimensions. In the first paper, we have studied the rotation curve of spiral galaxies; here, we have tackled the problem of the growth of cosmological structures. In the $\Lambda$CDM model, dark matter is essential in the growth of structures because it creates a potential well in which baryonic matter can collapse. Here, we have proved that, in principle, including extra dimensions, baryonic matter can collapse faster than in $\Lambda$CDM model even without dark matter. However, we still need a detailed numerical analysis on how the heat produced during the collapse is radiated away in our model, since if this process is inefficient, the collapse is slowed down or even halted. This is a matter for future works. We notice that Figures \ref{fig:contr} and \ref{fig:exp}, imply that we need contracting extra dimensions in order to have growth of structures in the early Universe. This is in line with what was found in \cite{mio}, where contracting extra dimensions were necessary to explain the accelerated expansion of the Universe even without dark energy. So we can say that the picture is consistent. The information on the number $n$ of extra dimensions must come from the comparison with observations, in particular with a fit of the luminosity distance to SNe data as we did in \cite{mio} or with the experiments we have proposed in \cite{mio2}.

It is known that redshift space distortions provide an independent estimate of the growth rate of the fluctuations, see \cite{S8,S85,S82}. They provide a measurement of $f\sigma_8$, where
\begin{equation}
    f=\dfrac{d\,ln(\delta)}{d\,\ln(a)}.
\end{equation}
If we use the $\delta$ derived from our analysis, we can predict the combination $f\sigma_8$ which is relevant to the study of the $S_8$ tension \cite{tension,S8,S85,S84}. Since our discussion gives a growth of structures different from that predicted $\Lambda$CDM (see \cite{S83}), it would be interesting to repeat the analysis of \cite{S8,S85,S84}, among the others, considering our predicted growth, in order to see if we could provide a way toward the solution of the $S_8$ tension.

The model presented in this paper and in \cite{mio,mio2,mio3} relies on the existence of extra dimensions, objects of which we do not have experimental evidence up to now; however, it is more economic since it gives a single explanation to very different phenomena (the rotation curve of spiral galaxies, the growth of structures and the accelerated expansion of the Universe) usually associated to obscure forms of matter or fields. This explanation is also well motivated theoretically in the works on Quantum Gravity, for example, by String Theory; therefore, we think that this model needs further attention.

\bibliography{biblio}

\end{document}